# Universal and Efficient p-Doping of Organic Semiconductors by Electrophilic Attack of Cations

*Jing Guo[1], Ying Liu[2], Ping-An Chen[1], Xinhao Wang[2], Yanpei Wang[2], Jing Guo[2], Xincan Qiu[1], Zebing Zeng[2], Lang Jiang[3], Yuanping Yi[3], Shun Watanabe[4], Lei Liao[1], Yugang Bai[2*], Thuc-Quyen Nguyen[5*], Yuanyuan Hu[1,6]\**

[1]Key Laboratory for Micro/Nano Optoelectronic Devices of Ministry of Education & Hunan Provincial Key Laboratory of Low-Dimensional Structural Physics and Devices, School of Physics and Electronics, Hunan University, Changsha 410082, China
[2]State Key Laboratory of Chem-/Bio-Sensing and Chemometrics, School of Chemistry and Chemical Engineering, Hunan University, Changsha, Hunan, 410082, China
[3]Beijing National Laboratory for Molecular Sciences, Key Laboratory of Organic Solids, Institute of Chemistry, Chinese Academy of Sciences, Beijing 100190, China
[4]Material Innovation Research Center (MIRC) and Department of Advanced Material Science, Graduate School of Frontier Sciences, The University of Tokyo, 5-1-5 Kashiwanoha, Kashiwa, Chiba, 77-8561, Japan
[5]Center for Polymers and Organic Solids, Department of Chemistry and Biochemistry, University of California at Santa Barbara, Santa Barbara, California 93106, United States
[6]Shenzhen Research Institute of Hunan University, Shenzhen 518063, China

Email of the corresponding authors:
baiyugang@hnu.edu.cn; quyen@chem.ucsb.edu; yhu@hnu.edu.cn

**ABSTRACT**

Doping is of great importance to tailor the electrical properties of semiconductors. However, the present doping methodologies for organic semiconductors (OSCs) are either inefficient or can only apply to a small number of OSCs, seriously limiting their general application. Herein, we reveal a novel p-doping mechanism by investigating the interactions between the dopant trityl tetrakis(pentafluorophenyl) borate (TrTPFB) and poly(3-hexylthiophene) (P3HT). It is found that electrophilic attack of the trityl cations on thiophenes results in the formation of alkylated thiophenium ions that induce electron transfer from neighboring P3HT chains, resulting in p-doping. This unique p-doping mechanism can be employed to dope various OSCs including those with high ionization energy (IE~5.8 eV). Moreover, this doping mechanism endows TrTPFB with strong doping ability, leading to polaron yielding efficiency of 100 % and doping efficiency of over 80 % in P3HT. The discovery and elucidation of this novel doping mechanism not only points out that strong electrophiles are a class of efficient p-dopants for OSCs, but also provides new opportunities towards highly efficient doping of OSCs.

# INTRODUCTION

Doping is an important method to modulate the electronic properties of semiconductors and is widely applied in semiconductor devices. The great success of inorganic silicon semiconductor industry is built on the basis of efficient, controllable and reliable doping of silicon. Interestingly, the emergence of organic electronics also starts from doping of polyacetylene (PA) with $I_2$ vapor, which leads to conductive polymer rather than insulator as reported by Heeger, MacDiarmid and Shirakawa in 1977[1], after which a new era of organic electronics based on organic semiconductors (OSCs) was launched. Doping has been proven to be essential for fabricating various high-performance organic devices including organic light-emitting diodes[2], organic photovoltaics (OPVs)[3], organic field-effect transistors (OFETs)[4] and organic thermoelectric generators (OTEGs)[5,6]. However, in contrast to doping in silicon, where a small amount of substitutional dopants can efficiently generate electrons (n-doping) or holes (p-doping), highly efficient and controllable doping of OSCs is an intensively studied topic yet remains a challenge[2, 7-9].

Taking p-doping of OSC as an example, presently there are mainly two mechanisms: redox reaction and protonation. The redox reaction doping mechanism requires matching the lowest unoccupied molecular orbital (LUMO, corresponds to electron affinity (EA)) level of the dopant with the highest occupied molecular orbital (HOMO, corresponds to ionization energy (IE)) level of the host semiconductor, by which integer or partial electron transfer can occur between them [8, 10-15]. This doping mechanism has been observed in molecular dopants such as $F_4TCNQ$, but it has several shortcomings or limitations. First, the mechanism fundamentally determines that only those OSCs with HOMO levels higher than the LUMO level of the dopant can be effectively doped, which is a crucial limitation to the general application of doping. As an example, OSCs with HOMO deeper than 5.3 eV cannot be well doped by $F_4TCNQ$ since the LUMO of $F_4TCNQ$ is 5.2 eV. Second, the doping efficiency (the ratio between the concentration of doping-induced free charge carriers and the concentration of dopants) based on this doping mechanism is rather low according to previous studies. For example, $F_4TCNQ$ generally shows a doping efficiency of about 15 % in $F_4TCNQ$:P3HT system[14, 16-19], although ultrahigh doping efficiency has been reported by adopting anion exchange or double doping strategies[8, 20]. Third, molecular dopants such as $F_4TCNQ$ have very low solubility in organic solvents (< 2 g $L^{-1}$ in THF) and tend to aggregate in thin films even at a low doping concentration[19, 21-25].

The other doping mechanism, *i.e.*, protonation doping, relies on the protonation of arenes on OSCs by strong Brønsted acids such as trifluoroacetic acid (TFA)[26] and HBr[27]. Recently, this doping mechanism was found to extend to Lewis acids such as tris(pentafluorophenyl)borane ($B(C_6F_5)_3$), which hydrolyzes to form Brønsted acids that can p-dope OSCs. Because the polarons are generated in spontaneous electron transfer processes initiated by the arenium ions, *i.e.*, protonated arenes, this doping mechanism places no constraints on the energy levels of host OSCs. For example, $B(C_6F_5)_3$ can p-dope several OSCs with IE larger than 5.2 eV[16, 17, 28]. However, the doping efficiency of $B(C_6F_5)_3$ is reported to be comparable to that of $F_4TCNQ$ when used as a dopant for P3HT, indicating the protonation doping mechanism does not promise sufficiently high doping performance[18]. This is very likely the result of high reversibility of arene protonation, which can significantly limit the efficiency of the electron transfer process that follows.

Recently, some novel dopants with remarkable doping performance have been reported[29-34]. For instance, we reported an organic salt trityl tetrakis(pentafluorophenyl) borate (TrTPFB) containing a trityl cation and the TPFB anion, which was shown to effectively p-dope organic semiconductors[35]. Wegner et al. reported a similar organic salt with borinium cation $Mes_2B$-TPFB ($Mes_2B^+$; Mes: mesitylene), and used this organic salt as a p-dopant in P3HT[33]. The borinium salt showed an occurrence of a bipolaronic peak, which indicates the strong doping ability of the salt. In addition to the superior doping performance, these organic salt dopants are found to have good miscibility with semiconductors, which is an important property desired for dopants. In spite of the promising applications of the organic salt dopants, their doping mechanism remains elusive.

In this work, we performed a series of experiments to understand the doping mechanism of TrTPFB, by which we reveal a new p-doping mechanism based on electrophilic attack of cations on OSCs. Specifically, we show that electrophilic attack of the trityl cation ($Ph_3C^+$) on thiophene rings leads to the alkylation and subsequent p-doping of P3HT. Importantly, such electrophilic attack behavior is expected to universally and efficiently happen between the trityl cation and OSCs, which explains the reason why the dopant is able to p-dope a wide range of OSCs regardless of their IE values, and to result in conductivity of up to 30 S cm$^{-1}$ in spin-coated P3HT films. Furthermore, in-depth investigations reveal that TrTPFB has nearly 100 % polaron yielding efficiency and a high doping efficiency of over 80 % in P3HT. These results suggest that the electrophilic attack doping mechanism has great potential for realizing highly efficient and universal doping in OSCs.

**RESULTS AND DISCUSSION**

**Universal p-doping effect of TrTPFB:** Figure 1a shows the molecular structures of TrTPFB and the polymer semiconductors we have used as host semiconductors (the full names of the semiconductors are shown in Methods section). The conductivity of all semiconductor films is observed to increase by solution-doping with TrTPFB (*i.e.* mixing the dopant with the OSCs in solution) regardless of the molecule structure or HOMO levels (Figure 1a). Specifically, the conductivity of P3HT at 10 mol% of TrTPFB doping concentration (corresponding to 10 dopants per 100 repeat units of 3-hexylthiophene) was improved by more than 6 orders of magnitude compared with the pristine P3HT film. The conductivity of several other semiconductors such as PCDTPT, PBDB-T, PTAA and PDVT-10, was improved by about 4 orders of magnitude; the conductivity of PBDB-T-SF and PBPTV was improved by about 3 orders of magnitude after doping. It is remarkable that the conductivity was enhanced by more than two orders of magnitude for N2200 and PFO with 30 mol% of TrTPFB, albeit their HOMO levels are around 5.8 eV. In fact, a p-channel field-effect transistor device based on N2200 can be realized by doping with TrTPFB, which clearly shows the p-doping effect (see Supporting Information S1.3).

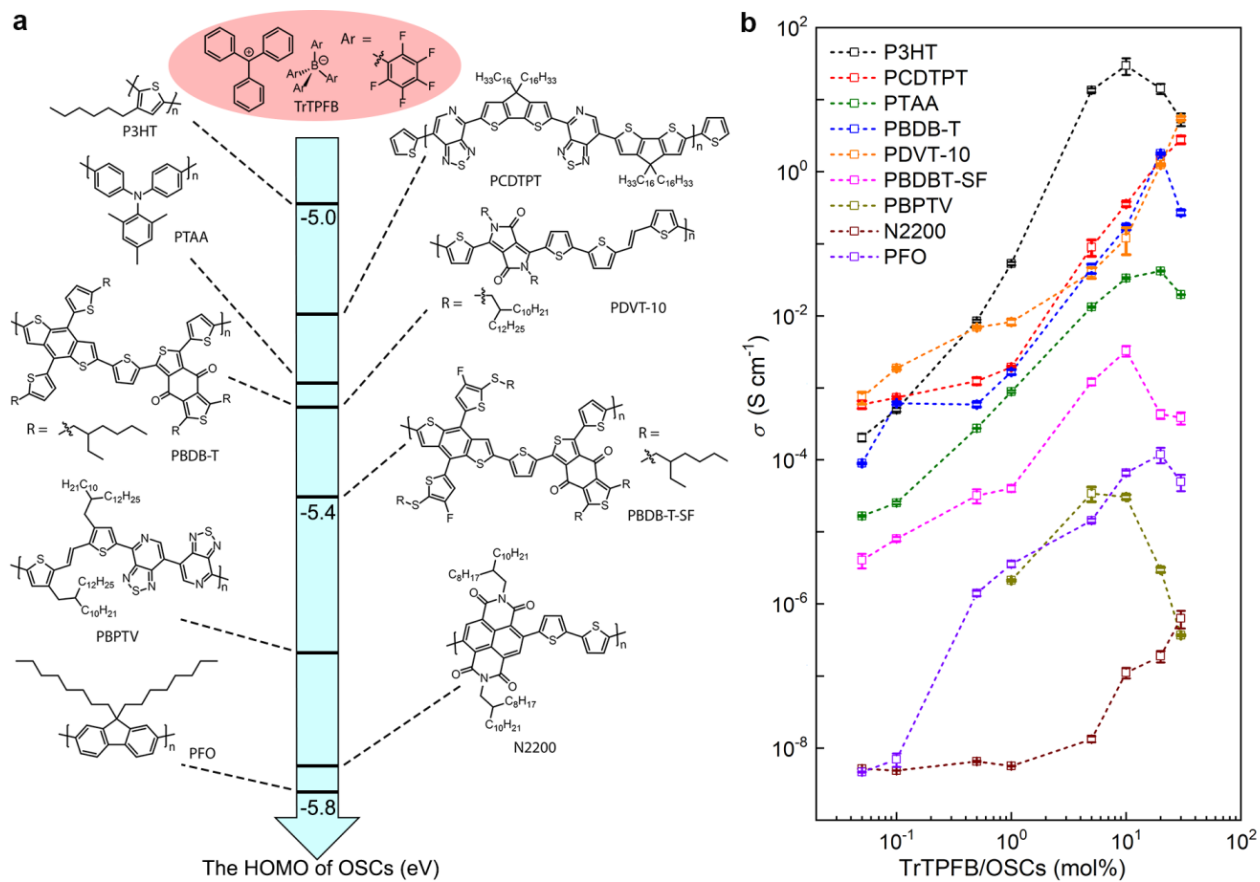

**Figure 1. The universal p-doping effect of TrTPFB.** (a) The molecule structure of TrTPFB and OSCs used in this study, with their HOMO levels illustrated. (b) The electrical conductivity of a series of OSCs doped with TrTPFB as a function of doping concentration.

**Doping mechanism of TrTPFB:** The universal p-doping effect of TrTPFB shown above is undoubtedly an attractive property for dopants, and inspires us to understand the doping mechanism. Indeed, the LUMO and HOMO of TrTPFB were estimated to be −4.61 eV and −6.98 eV, respectively, by performing cyclic voltammetry (CV) and UV-vis-NIR measurements (Supporting Information S3). Such energy levels of TrTPFB make it thermodynamically difficult to p-dope

P3HT through direct redox chemical doping, thus impelling us to propose a different doping mechanism for explaining the doping effect. Enlightened by the protonation-doping mechanism (Figure 2a, top), we propose the doping scheme using TrTPFB as the dopant for P3HT. In nature, TrTPFB is a salt with a sterically hindered carbenium ion, *i.e.*, the trityl cation $Ph_3C^+$. Carbenium ions are strong electrophiles as they are commonly seen in one of the earliest name reactions: Friedel-Crafts reaction[36]. Thus, it is not surprising to see electrophilic attack of the trityl cation on thiophene, which is an electron-rich arene. Importantly, this electrophilic attack by $Ph_3C^+$ would not lead to the Friedel-Crafts reaction in P3HT but result in the alkylated arenium (thiophenium) ion, alternatively called the Wheland intermediate (Figure 2a, bottom), which is structurally similar to the protonated thiophene but more stable because of the sterically hindered substituent ($Ph_3C$) (see more details Supporting Information S5)[37]. The existence of this Wheland intermediate can induce electron transfer that is responsible for p-doping[28, 38]. Such p-doping processes are illustrated in Figure 2b.

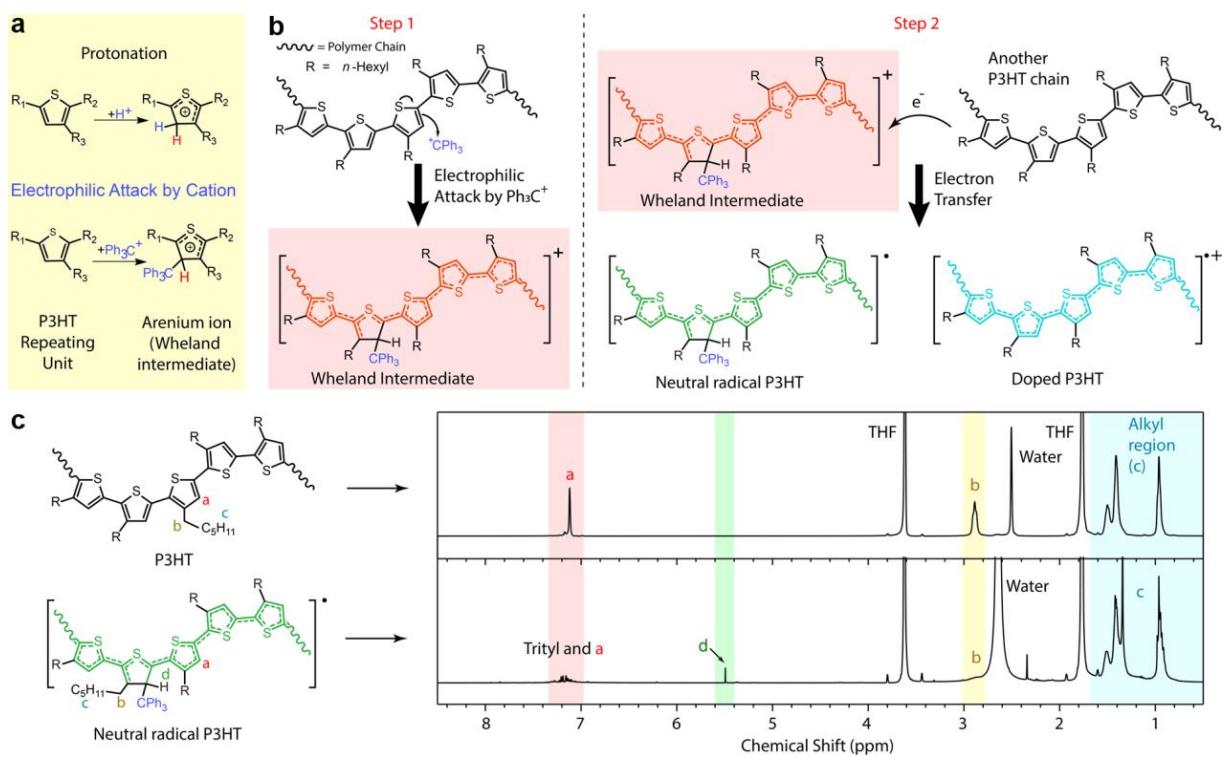

**Figure 2. Schematic illustration and evidences for the electrophilic-attack-based p-doping mechanism.** (a) Schematic illustration of the protonation of thiophene involved in the protonation-based doping mechanism, and of the electrophilic-attack-based alkylation reaction that yields stabilized Wheland intermediate. (b) Proposed p-doping mechanism of TrTPFB on P3HT in steps. The electrophilic attack of trityl cation on the thiophene unit yields thiophenium ion (the Wheland intermediate), which subsequently oxidizes another thiophene unit to achieve p-doping. (c) $^1$H

NMR characterization of P3HT and TrTPFB-doped P3HT. Successful alkylation (as suggested by the peak of the proton colored in green) and formation of radicals are clearly observed.

To verify if the doping proceeds in the way we postulated, we characterized the precipitates produced when P3HT was doped with TrTPFB at a high doping ratio in deuterated tetrahydrofuran (THF-$d_8$) using $^1$H NMR (See supporting information S4.1 and S4.2), and a new peak at 5.5 ppm was observed after doping (peak d in Figure 2c). This new peak was assigned to the non-aromatic proton on the thiophene unit formed upon electrophilic attack of trityl cation[28]. Furthermore, all signals for the protons near the large conjugated system almost disappeared (peak b in Figure 2c), indicating the formation of radicals that move along the connected π bonds, as radicals can significantly shorten $T_2$ relaxation time of nearby nuclei and make their peaks highly broadened and disappearing[39]. Thus, these NMR spectroscopic evidence well supported the formation of the tritylated thiophenium and the subsequent electron transfer from a native P3HT chain to an tritylated P3HT chain which leads to polaron formation in the semiconductors (Figure 2b)[28]. Moreover, results from a model study using a thiophene tetramer (OT$_4$) further supported our hypothesis on the covalent linking of the trityl groups to thiophene units through electrophilic attack and the assignment of peak d (see Supporting Information 4.3).

This electrophilic-attack-based doping mechanism owns two prominent advantages: first, since all OSCs contain aromatic units that can be the target of an electrophilic attack, p-doping through this electrophilic attack mechanism is supposed to happen in all OSCs regardless of their energetic levels, which is responsible for the universal doping effect as shown in Figure 1b. Second, in contrast to the arenium ion from the protonation of thiophene, which is usually an unstable species subject to fast re-aromatization through deprotonation, the arenium ion formed through the alkylation of thiophene by a bulky trityl cation we report here has assured stability. In specific, by forcing the arenium ring to adopt a conformation unfavorable for the dissociation of proton (C-H bond at equatorial position, see Supporting Information S5.1 and S5.2), the deprotonation and re-aromatization of the alkylated thiophenium is kinetically unfavorable[37]; the stabilized arenium system thus promises the achievement of high doping efficiency, which will be further illustrated below. More importantly, this doping mechanism is not only applicable to trityl cation, but also to other strong electrophiles, such as nitronium tetrafluoroborate and diphenyliodium tetrakis(pentafluorophenyl)borate, which were also observed to produce significant doping effect when they were mixed with P3HT (Supporting Information S5.3). Presumably, these compounds also undergo electrophilic attack on thiophene, generating arenium ions responsible for effective doping. Therefore, our new doping approach is a generally applicable one that can be extended to numerous dopants.

**Ultrahigh doping efficiency of TrTPFB:** In this part, we performed quantitative studies to investigate the doping performance of TrTPFB to gain insight into the superiority of the electrophilic-attack-doping mechanism. The strong doping ability of TrTPFB can be apparently seen from the considerably higher conductivity of TrTPFB-doped P3HT than that of P3HT doped by a reference dopant $F_4TCNQ$ (Figure 3a). The maximum conductivity of the P3HT film reaches 30 S cm$^{-1}$ at 10 mol% of TrTPFB doping concentration, which is indeed a record-high value for spin-coated P3HT films processed by solution-doping method, as shown in Figure 3b. In fact, the superior doping performance of TrTPFB to $F_4TCNQ$ is not only observed in P3HT, but also in other polymer semiconductors (see Supporting Information S1.4).

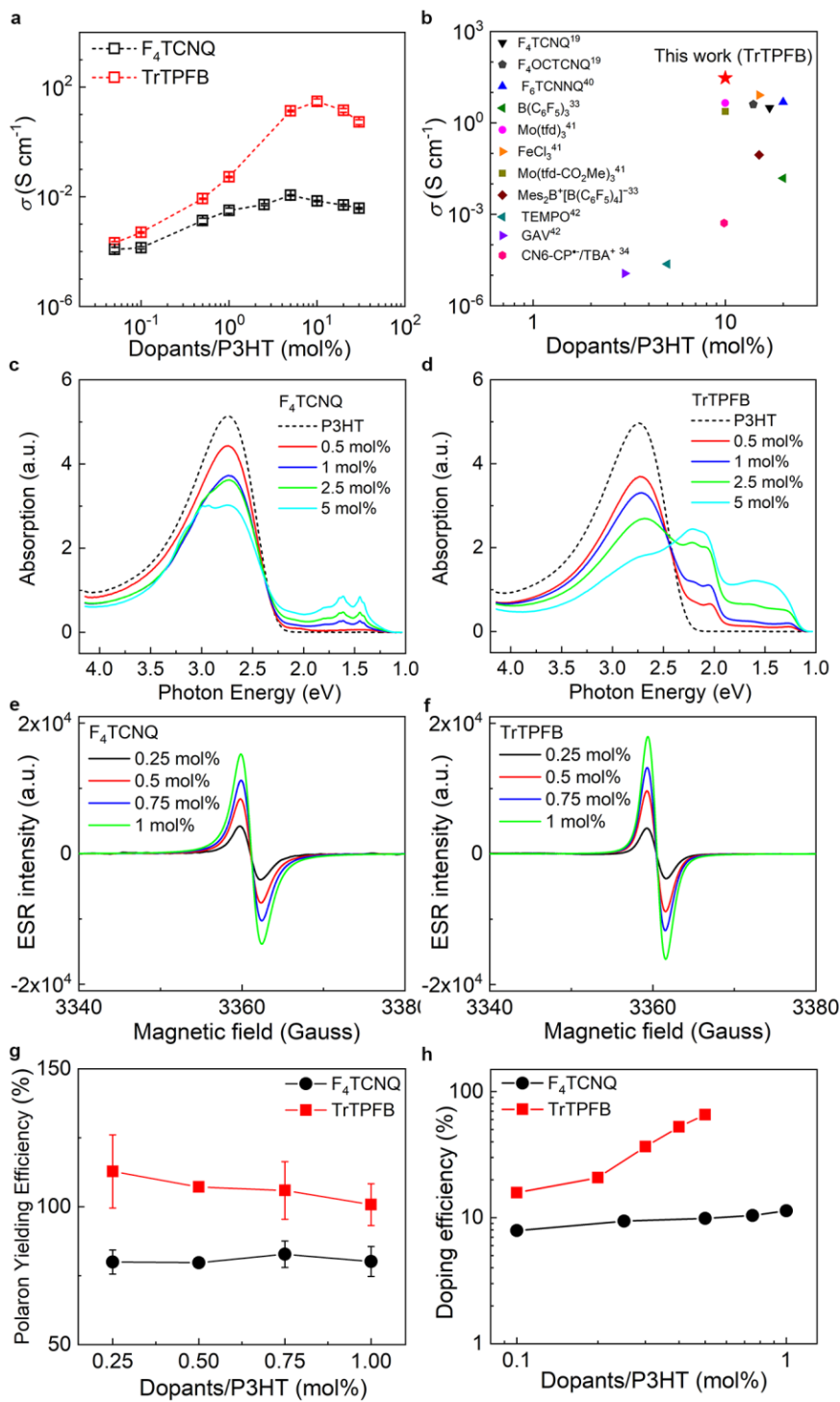

**Figure 3. Ultrahigh doping efficiency of TrTPFB.** (a) The electrical conductivity of P3HT doped with TrTPFB and F$_4$TCNQ as a function of doping concentration. (b) The conductivity of spin-coated P3HT films doped by TrTPFB and other dopants from references[19, 33, 34, 40-42] using the solution doping method[41-43]. Optical absorption spectra of doped P3HT solutions with (c) F$_4$TCNQ as dopant and (d) TrTPFB as dopant. ESR signals of (e) F$_4$TCNQ- and (f) TrTPFB-doped P3HT films at different doping ratios. (g) The extracted polaron yielding efficiency for the two dopants. (h) The doping efficiency extracted from the Mott-Schottky characterizations.

We then employed optical absorption spectroscopy to characterize the dopant-induced polarons[12, 44]. Since F$_4$TCNQ has been well studied as a dopant for P3HT, it was taken as a reference dopant for the study[12, 45, 46]. The absorption spectra of F$_4$TCNQ- and TrTPFB-doped P3HT solutions as a function of doping ratio are shown in Figure 3c and d. A broad absorption in the range of 1.1 to 1.75 eV is observed upon doping in both solutions, which is ascribed to the dopant-induced polaron on the P3HT backbone[44, 45]. Aside from these absorptions, there also exists a broad absorption peak around 2.2 eV, which was assigned as the absorption of neutral ordered P3HT aggregates[46, 47]. Previous studies show that the absorption peaks of P3HT polarons are located at about 1.4 and 1.6 eV[18, 33, 44, 45], based on which we can compare the doping magnitude of the two by comparing their polaron absorption intensity. It is seen that the polaron absorption for TrTPFB at 1.6 eV is about 1.5 times higher than that for F$_4$TCNQ at the same doping ratio (Supporting Information S6.1), implying that more polarons are produced in TrTPFB-doped P3HT (See Supporting Information S6.1).

We further employed electron spin resonance (ESR) technique to characterize the polaron numbers and to estimate the polaron yielding efficiency, which is defined as the ratio between the number of polarons and the number of dopants, in the two doped films. The number of polarons at different doping concentrations can be quantitatively extracted from the ESR signals (Figure 3g and more details in Supporting Information S6.2). It is found that the polaron yielding efficiency of F$_4$TCNQ is about 70 % in the doping range of 0.25 mol% to 1 mol%, which is consistent with previous results[12, 18, 45]. In comparison, the polaron yielding efficiency of TrTPFB is as high as 100 %, indicating that each dopant (trityl cation) can induce a polaron in P3HT, which is impressive and, to the best of our knowledge, is for the first time reported for dopants in P3HT.

However, it is notable that the production of polarons is the first key step in the doping process, which results in ion pairs (polaron and dopant anion) that are bound by Coulombic force. Following that ion pairs need to overcome Coulombic binding by thermally dissociating into separate charge carriers[48]. Consequently, the doping efficiency, which is defined as the ratio of number of free charge carriers to number of dopant molecules, is determined by the efficiency of the two processes and is generally lower than the polaron yielding efficiency. To estimate the doping efficiency of the two dopants, Mott-Schottky analysis was performed on metal-insulator-semiconductor (MIS) diodes for extraction of dopant-induced charge carrier density[49] (See Supporting Information S6.3). The doping efficiency of the two dopants is shown in Figure 3h. F$_4$TCNQ was found to have doping efficiency of about 10 % in the doping range of 0.1-1 mol%, which is consistent with previous reports[18]. In contrast, the doping efficiency of TrTPFB is around 20 % at low doping concentration, possibly due to the trap-filling of charge carriers, while this value quicky increases to 80 % at the doping ratio of 0.5 mol%, which, to the best of our knowledge, is among the highest doping efficiency values ever reported for solution-doped P3HT[20]. In particular, due to the super linear increase of charge carrier density with doping ratios as seen in Figure 3h, it is very likely that the doping efficiency of TrTPFB for P3HT can be further higher as doping ratios increases. Such results unambiguously demonstrate the excellent doping ability of TrTPFB as a p-dopant.

In view of the high doping performance of TrTFFB on P3HT, we evaluated the thermoelectric performance of this system by measuring the Seebeck coefficients (*S*) of doped P3HT films (see supporting information S7). The power factor (*PF*), which is an important figure of merit for characterizing thermoelectric performance, was obtained by equation: $PF = S^2\sigma$. The maximum *PF* of P3HT films doped by TrTPFB reaches 27.8 μW m$^{-1}$K$^{-2}$. This *PF* value is also the highest one for spin-coated P3HT films processed by solution doping method, as shown in Figure S24. All these results directly illustrate the outstanding doping ability of TrTPFB.

**The influence of anion structure on doing performance:** Having identified the doping mechanism of trityl cation and its doping performance, it is intriguing to know how the [B(C$_6$F$_5$)$_4$]$^-$ anion affects the doping process. Thus, we have carried out preliminary investigations by characterizing the doping performance of two other organic salts consisting of trityl cation but different anions: trityl tetrafluoroborate (TrBF$_4$) and trityl hexafluophosphate (TrPF$_6$). Figure 4a shows the ESR spectra of P3HT films doped by the two organic salts, which apparently illustrates their doping effect. Meanwhile, it is noted that the conductivity of P3HT doped by these two dopants is much lower compared to the one doped by TrTPFB at the same dopant concentration (see Figure 4b), implying the weaker doping performance of these two dopants.

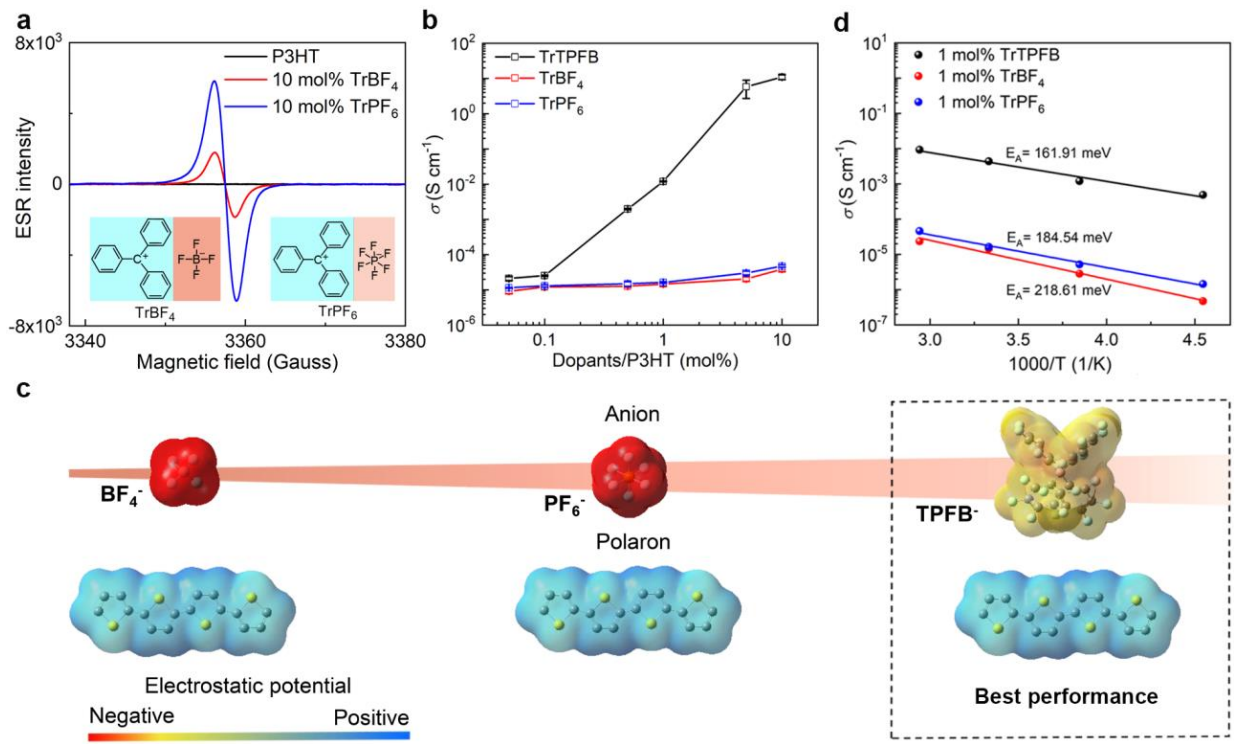

**Figure 4. Influence of anion structure on doping performance.** (a) The ESR spectra of P3HT films doped with TrBF$_4$ and TrPF$_6$. The molecular structures of two organic salts are shown in the inset. (b) The conductivity of doped P3HT as a function of doping concentration. (c) Schematic diagram showing the dependence of cation-anion and polaron-anion Coulombic bounding on the anion sizes. The electrostatic surface potential (ESP) of all cations and anions is determined by density of functional theory (DFT) calculations. (d) The temperature-dependent conductivity of P3HT doped with the three dopants.

These results suggest the molecule structure of the anion definitely affects the doping efficiency and thus should be properly chosen. Our results show that [B(C$_6$F$_5$)$_4$]$^-$ is preferred over BF$_4^-$ and PF$_6^-$ for achieving higher doping efficiency. This is probably accounted by several reasons: (1) the delocalization of negative charge over a large anion such as [B(C$_6$F$_5$)$_4$]$^-$ helps with the stabilization of the anion and more importantly, could be favorable for the electrophilic attack reaction shown above, as revealed by Beljonne *et al.* through theoretical calculations[38]. (2) the bulky structure of [B(C$_6$F$_5$)$_4$]$^-$ leads to low negative electrostatic surface potential and reduces the electrostatic attractive force between it and the polaron, which is beneficial to efficient ion pair dissociation or a higher charge dissociation efficiency[41, 50, 51] (see Figure 4c). Indeed, the doped P3HT films were seen to exhibit increasing activation energy with the decrease of the anion size according to temperature dependent conductivity measurements (Figure 4d), which is partly attributed to the increasing electrostatic binding energies between the generated polarons and the anions. Besides the ability to offer high doping efficiency, the $[B(C_6F_5)_4]^-$ was also reported to be an inert non-coordinating counterion that is moisture- and air-resistant, which is important to stable doping[33]. The results shown in Supporting Information S8 demonstrate the high storage stability of TrTPFB-doped P3HT films.

**CONCLUSION**

To summarize, we reveal a new p-doping mechanism relying on the electrophilic attack of cations on OSCs by investigating the interaction between TrTPFB and P3HT. This doping mechanism intrinsically possesses significant advantages over the previously reported doping mechanisms, and endows the corresponding dopants with ability to universally and efficiently p-dope OSCs. We show that TrTPFB exhibits superior doping performance with the ability to dope various OSCs even if the HOMO level of the semiconductor is as high as 5.8 eV, and to enhance the conductivity of P3HT to 30 S cm$^{-1}$. In-depth investigations indicate the polaron yielding efficiency and doping efficiency of TrTPFB reach ultrahigh values of 100 % and 80 %, respectively. Besides, we find that although the anion is not directly involved in the doping reaction, its structure greatly affects the doping efficiency and thus should be properly chosen. Our study casts light on the doping physics of OSCs, and provides vast opportunities to realize efficient and controllable doping of OSCs.

**Methods**

*Materials:* Poly[4-(4,4-dihexadecyl-4H-cyclopenta[1,2-b:5,4-b']-dithiophen-2-yl)-alt-[1,2,5]thiadiazolo-[3,4-c]pyridine] (PCDTPT), Poly[[4,8-bis[5-(2-ethylhexyl)-2-thienyl]benzo[1,2-b:4,5-b']dithiophene-2,6-diyl]-2,5-thiophenediyl[5,7-bis(2-ethylhexyl)-4,8-dioxo-4H,8H-benzo[1,2-c:4,5-c'] dithiophene-1,3-diyl]] (PBDB-T), Poly[bis(4-phenyl)(2,4,6-trimethylphenyl)amine (PTAA), Poly{3,6-dithiophen-2-yl-2,5-di(2-decyltetradecyl)-pyrrolo[3,4-c]pyrrole-1,4-dione-alt-thienylenevinylene-2,5-yl} (PDVT-10), Poly[(2,6-(4,8-bis(5-(2-ethylhexylthio)-4-fluorothiophen-2-yl)-benzo[1,2-b:4,5-b']dithiophene))-alt-(5,5-(1',3'-di-2-thienyl-5',7'-bis(2-ethylhexyl)benzo[1',2'-c:4',5'-c']dithiophene-4,8-dione)] (PBDB-T-SF), Pyridal[2,1,3]thiadiazole-based semiconducting polymer (PBPTV), Poly{[N,N'-bis(2-octyldodecyl)naphthalene-1,4,5,8-bis(dicarboximide)-2,6-diyl]-alt-5,5'-(2,2'-bithiophene)} (N2200) and Polydioctylfluorene (PFO) were purchased from commercial companies and used as received without further purifications. P3HT were purchased from Aladdin Reagent Co., Ltd.

*Preparation of doped solutions and films:* TrTPFB (from Strem Chemicals, Inc.), F$_4$TCNQ, TrPF$_6$, TrBF$_4$ (from TCI (Shanghai) Development Co., Ltd) and all OSCs were dissolved in chlorobenzene to prepare semiconductor solutions with different dopant concentrations unless specified with other solvents. The concentration of P3HT solution was 20 g L$^{-1}$, while the solution concentrations of TrTPFB (0.1 g L$^{-1}$, 0.5 g L$^{-1}$, 1 g L$^{-1}$, 5 g L$^{-1}$ and 10 g L$^{-1}$), F$_4$TCNQ (0.1 g L$^{-1}$ and 0.5 g L$^{-1}$), TrPF$_6$ (0.1 g L$^{-1}$ and 0.5 g L$^{-1}$) and TrBF$_4$ (0.1 g L$^{-1}$ and 0.5 g L$^{-1}$) varied with the doping ratio. The blended solutions were stirred to be homogeneous prior to film preparation. All solutions were filtered through the 0.45 µm syringe filters to remove impurities or aggregates before usage. Semiconductor films were made through spin-coating of solutions in glovebox, and different post-annealing conditions were used for different semiconductors (Supporting Information S1)

*CV, NMR and MS measurements:* For Cyclic voltammogram (CV) of TrTPFB solution in dichloromethane (1.0 mM), redox potentials were determined by using 0.10 M n-Bu$_4$N$^+$PF$_6^-$ as a supporting electrolyte, and the electrode potential was externally calibrated by the ferrocene / ferrocenium redox couple. $^1$H NMR and HMBC spectra were recorded on a Bruker AVANCE Neo 400 spectrometer. Typically, the OT$_4$ CDCl$_3$ solutions and TrTPFB CDCl$_3$ solutions were mixed at 100 mol% and stored in the glovebox for 24 hours before $^1$H NMR and HMBC measurement at room temperature. The precipitate produced by TrTPFB doped P3HT was dissolved in THF-$d_8$ and recorded at 60°C. High resolution ESI-MS characterization were carried out with a Bruker compact quadrupole time-of-flight mass spectrometry system (QTOF-MS). The doped OT$_4$ solutions were dispersed in methanol and filtered before the measurement.

*UV-vis-NIR and ESR measurements:* The ultraviolet-visible-near-infrared (UV-vis-NIR) absorption spectra of solution samples were carried out with UV-3600PLUS (SHIMADZU). For ESR measurements, the prepared sample solutions were dropped onto glass substrates and dried in an Ar-filled glove box to remove the solvent, and then placed into paramagnetic tubes. After sealing the paramagnetic tubes, the ESR was measured on a JEOL JES-FA200 ESR spectrometer at room temperature.

*Measurement of electrical conductivity:* Patterned electrodes (Cr/Au: 2 nm/30 nm) were prepared on Si wafers with 300 nm SiO$_2$ by photolithography. The doped semiconductor solutions were spin-coated to form a thin film on the substrate at a speed of 1500 rpm for 30 s and annealed at the corresponding temperature and time (see Supporting Information Table S1 in detail). The conductivity of the prepared devices was measured in air by a four-probe method through a Keithley 4200 semiconductor analyzer.

*Preparation and measurement of OFETs and MIS diodes:* In this work, top-gate, bottom-contact OFETs were fabricated. Firstly, patterned electrodes (Cr/Au: 2 nm/30 nm) were prepared on a glass substrate by photolithography. Secondly, the substrate was sonicated for 5 min in ultrapure water, acetone and isopropanol in sequence. Thirdly, the prepared dopant/OSC solutions were spin-coated on the substrate at 1500 rpm for 30 s, and then they were annealed according to the corresponding conditions (see Supporting Information Table S1). Subsequently, Cytop solution was spin-coated at 1000 rpm for 30 s and baked at 90 °C for 20 min as a dielectric layer. Finally, Al with a thickness of about 100 nm was deposited as a gate electrode through a shadow mask. The OFETs were measured in an Ar-filled glove box by an Angilent B2912A source meter.

MIS Diodes were prepared by spin coating the prepared doping solutions on ITO glass substrates with about 30 nm $Al_2O_3$ deposited by ALD Atomic Layer Deposition (ALD), following which about 40 nm of Au was evaporated. Next, the capacitance-voltage measurement was performed on Agilent 4294A precision impedance analyzer at frequency of 100 Hz, AC voltage of 0.5 V, with DC bias voltage in the range of 10 V to −10 V.

*Characterization of thermoelectric performance:* To measure the Seebeck coefficient of the doped samples, a home-made thermoelectric measurement setup was used. The devices containing one heater, two thermometers which also act as electrical contacts were fabricated by photolithographic patterning of metal bilayers of Cr (10 nm) and Au (15 nm) on glass substrates. To obtain Seebeck coefficient $S = \frac{\Delta V}{\Delta T}$, the temperature gradient between the two electrodes was estimated by converting the resistance of electrodes into temperature using the temperature-coefficient-of resistance (TCR) (see supporting information S7), and the built-in thermal voltage was measured using Keithley nanovoltmeter model 2182A. All Seebeck coefficients were measured at 300 K in high vacuum (<$10^{-5}$ mbar) using Janis ST-100.

*ESP calculation*: DFT calculations were performed with the Gaussian 09 program. All geometry optimizations were carried out at the B3LYP level of DFT with the 6-31G (d, p) basis set. ESP of five compounds were calculated at B3LYP/6-31G (d, p) level.

## AUTHOR INFORMATION

**Corresponding Author**

E-mail: baiyugang@hnu.edu.cn; quyen@chem.ucsb.edu; yhu@hnu.edu.cn

**Author contributions**

J.G. performed the characterizations of electrical conductivity as well as the absorption spectra of different organic semiconductors doped by different dopants. Y.L. and Y.B. performed NMR, MS and FT-IR analyses. X.W., Y.W., J.G. and Z.Z. provided help with absorption spectra and ESR measurements. J.G., P.C., X.Q. performed the fabrication and measurement of MIS diodes. P.C. performed measurement of the Seebeck Coefficient. H.C. provided polymer semiconductor materials. L. L. provided help with electrical measurements. Y.B., T.-Q. N and Y.H. conceived the idea and supervised the project. J.G., Y.L., L.J., S. W., Y.B., T.-Q. N and Y.H. wrote the manuscript. All the authors revised and approved the manuscript.

**Notes**

The authors declare no competing financial interests.

## ACKNOWLEDGMENT

We thank Dr. Tong Xia from Central South University for help with the decomposition of UV-vis-NIR spectrum, and Prof. Jianpu Wang from Nanjing Tech University for helpful discussions. The work is supported by the National Key Research and Development Program (2021YFA1200700), National Natural Science Foundation of China (Grant No. 62074054; 21877033, 92163127), the Natural Science Foundation of Hunan Province (Grant No. 2020JJ1002). T.-Q.N. acknowledges the support from the US Department of Energy under Award no. DE-SC0017659.